\documentstyle[aps,prl,amssymb,latexsym,epsfig]{revtex}

\newcommand{\beq}{\begin{equation}}
\newcommand{\eeq}{\end{equation}}
\newcommand{\bqa}{\begin{eqnarray}}
\newcommand{\eqa}{\end{eqnarray}}
\newcommand{\fr}{\frac}

\begin{document}

\draft

\wideabs{
\title{Naked singularities in Tolman-Bondi-de Sitter collapse}
\author{S\' ergio M. C. V. Gon\c calves}
\address{Theoretical Astrophysics, California Institute of Technology, Pasadena, California 91125}
\date{October 6, 2000}
\maketitle
\begin{abstract}
We study the formation of central naked singularities in spherical dust collapse with a cosmological constant. We find that the central curvature singularity is locally naked, Tipler strong, and generic, in the sense that it forms from a non-zero-measure set of regular initial data. We also find that the Weyl and Ricci curvature scalars diverge at the singularity, with the former dominating over the latter, thereby signaling the non-local origin of the singularity.
\end{abstract}
\pacs{PACS numbers: 04.20.Dw, 04.20.Jb, 04.70.Bw}
}

\narrowtext

\section{INTRODUCTION}

One of the outstanding issues in general relativity is that of the final state of gravitational collapse, a crucial aspect of which is the possibility of spacetime singularities---events at which Riemannian curvature typically diverges, the spacetime is geodesically incomplete, and any classical theory of gravity necessarily breaks down. It has long been known that under a variety of circumstances, spacetimes which are solutions of Einstein's equations with physically reasonable regular initial data, inevitably develop singularities \cite{hawking&ellis73}. 

In an effort to protect the applicability of general relativity, Penrose conjectured that such singularities must be hidden by an event horizon, and thus invisible to an asymptotic observer, i.e., they cannot be {\em globally} naked \cite{penrose69}. This constitutes in essence what has become known as the {\em weak cosmic censorship conjecture.} However, it is quite possible---at least in principle---for an observer to penetrate the event horizon and live a rather normal life inside a black hole. This motivated the {\em strong cosmic censorship conjecture}, which broadly states that timelike singularities cannot occur in nature, i.e., there are no {\em locally} naked singularities \cite{penrose79}.

A lack of tools to handle global properties of the Einstein equations (and respective solutions), together with their high non-linearity, have been the main obstacle to provable formulations of the (weak or strong) cosmic censorship conjecture. Whilst efforts are being undertaken in this direction \cite{global}, one can hope that the detailed study of specific models helps to isolate some defining features of singularity formation and structure, thereby contributing towards a precise, counter-example-free formulation of the conjecture.

One such model is inhomogeneous Tolman-Bondi collapse \cite{tolman34,bondi48}, whose general solution is analytically obtainable in closed (or parametric) form. For this reason, and because spherical symmetry is, arguably, a plausible approximation for the geometry at the late stages of collapse \cite{nakamura&sato82}, Tolman-Bondi collapse has been extensively studied by many authors \cite{tbauthors2,tbauthors}. The main results stemming from these analyses, pertinent to the context of singularities and cosmic censorship, may be summarized as follows: (i) A central curvature singularity always forms; (ii) This singularity can be either locally or globally naked, depending on the initial data; (iii) The singularity is generic, in the sense that an infinite number of outgoing non-spacelike geodesics terminates at the singularity in the past; (iv) The singularity is gravitationally strong in the sense of Tipler \cite{tipler77}. 

We remark that inhomogeneity plays a key role in the global visibility of the singularity in asymptotically flat spherical dust collapse, with the former being uniquely determined by the derivatives of the initial central energy density profile. In connection with the cosmic censorship conjecture and the physical meaningfulness of the singularity, it should be noted that the central naked singularity of inhomogeneous Tolman-Bondi collapse is always Tipler strong, irrespective of the initial data \cite{deshingkar&joshi&dwivedi99}, and, furthermore, it is marginally stable against non-spherical linear perturbations \cite{hin98-00}. 

Recent observations of high-redshift Type Ia supernovae \cite{riess98,perlmutter98} and peculiar motion of low-redshift galaxies \cite{zehani&dekel99}, appear to indicate that the present radius of the universe is accelerating, thus suggesting the existence of a positive cosmological constant, $\Lambda>0$. This has sparked a renewed interest in gravitational collapse with a cosmological constant \cite{omer65}. Markovic and Shapiro \cite{markovic&shapiro00} analyzed the spherical homogeneous collapse of a dust cloud in the presence of a positive cosmological constant, and found that depending on the initial data (cosmological constant, gravitational mass of the cloud, and its comoving radius), the resulting spacetime can be either (i) a Schwarzschild-de Sitter black hole \cite{kottler18}, (ii) a bouncing sphere, or (iii) a de Sitter-like global cosmological singularity. Their analysis was qualitatively generalized to the inhomogeneous and degenerate cases, for both $\Lambda>0$ and $\Lambda<0$, by Lake \cite{lake00}. The collapse of null dust with a negative cosmological constant was studied by Lemos \cite{lemos99}, who showed that a Tipler strong, globally naked singularity develops for spherical collapse (but not for toroidal, cylindrical, or planar geometries).

In this paper, we examine in detail the effects of a positive cosmological constant in the singularity formation and structure in spherical dust collapse in an asymptotically de Sitter spacetime.  The existence of $\Lambda>0$ changes the nature of the solution not only at large radii, but also near the center, and thus the central singularity---if it exists---may have a different structure from that occurring in asymptotically flat spherical dust collapse. The current status of the cosmological constant makes pertinent the analysis of its effects on singularity formation and structure, particularly in the cosmic censorship context. Does a central curvature singularity exist in spherical dust collapse with $\Lambda>0$? If it does, is it naked and gravitationally strong? As we shall show below, both questions are answered positively.

For definiteness, we shall hereafter refer to spherical dust solutions of the Einstein equations with a cosmological constant as Tolman-Bondi-de Sitter spacetimes.

The paper is organized as follows: Section II derives the Tolman-Bondi-de Sitter metrics from the Einstein equations. In Sec. III, the existence and genericity of a central curvature singularity are proven. Section IV discusses the singularity's curvature strength and the relative strength of the Weyl and Ricci scalars. Section V concludes with a summary and discussion.

Geometrized units, in which $G=c=1$, are used throughout.

\section{TOLMAN-BONDI-DE SITTER SPACETIMES}

The Tolman-Bondi-de Sitter family of solutions is given by a spherically symmetric metric, written here in normal Gaussian coordinates $\{t,r,\theta,\phi\}$:
\bqa
ds^{2}&=&-dt^{2}+e^{-2\Psi(t,r)}dr^{2}+R^{2}(t,r)d\Omega^{2}, \label{metb}\\
d\Omega&\equiv&d\theta^{2}+\sin^{2}\theta d\phi^{2},
\eqa
together with the stress-energy tensor:
\beq
T_{ab}=\rho(t,r)u_{a}u_{b}-\fr{\Lambda}{8\pi}g_{ab}=\rho\delta_{a}^{t}\delta_{b}^{t}-\fr{\Lambda}{8\pi}g_{ab},
\eeq
where $u^{a}=\delta^{a}_{t}$ is the 4-velocity of a dust element, $\rho(t,r)$ is the energy density, and $\Lambda$ the cosmological constant.

With the metric (\ref{metb}), the independent non-vanishing Einstein tensor components are
\bqa
G_{tt}&=&R^{-2}[-Re^{2\Psi}(2R'\Psi'+2R''+R^{-1}R'^{2}) \nonumber \\
&&-2\dot{R}\dot{\Psi}R+1+\dot{R}^{2}], \\
G_{rt}&=&-2R^{-1}(\dot{R}'+R'\dot{\Psi}), \\
G_{rr}&=&-R^{-2}\left[e^{-2\Psi}(2\ddot{R}R+\dot{R}^{2}+1)-R'^{2}\right], \\
G_{\theta\theta}&=&\sin^{-2}\theta\,G_{\phi\phi}=R(\dot{R}\dot{\Psi}+R'\Psi'e^{2\Psi} \nonumber \\
&&+R''e^{2\Psi}-\ddot{R}+\ddot{\Psi}R-\dot{\Psi}^{2}R),
\eqa
where the overdot and prime denote partial differentiation with respect to $t$ and $r$, respectively.

Introducing the auxiliary functions
\bqa
k(t,r)&\equiv&1-e^{2\Psi}R'^{2}, \label{kappa} \\
m(t,r)&\equiv&\fr{1}{2}R\left(\dot{R}^{2}+k-\fr{\Lambda}{3}R^{2}\right), \label{masstb}
\eqa
Einstein's equations\footnote{Since there are only three functions to be determined and four equations, only three of these are independent, with the remaining one acting as a constraint. We take Eqs. (\ref{tb1})-(\ref{tb3}) as our complete set, and Eq. (\ref{mtb}) as the constraint equation, since it provides a simple relation between the initial data and the initial mass profile.} simplify greatly to
\bqa
\dot{R}^{2}&=&2mR^{-1}-k+\fr{\Lambda}{3}R^{2}, \label{tb1} \\
\dot{k}&=&0, \label{tb2} \\
\dot{m}&=&0, \label{tb3}
\eqa
with the constraint
\beq
m'=4\pi R^{2}R'\rho(t,r). \label{mtb}
\eeq

The Tolman-Bondi-de Sitter metrics are then given by
\beq
ds^{2}=-dt^{2}+\fr{R'^{2}}{1-k}dr^{2}+R^{2}d\Omega^{2}, \label{tbm}
\eeq
where $R(t,r)$ is a solution of Eq. (\ref{tb1}), with initial data given by Eq. (\ref{mtb}),
\beq
m(r)=4\pi\int_{0}^{r} R^{2}(0,\tilde{r})R'(0,\tilde{r})\rho(0,\tilde{r})d\tilde{r}, \label{mass}
\eeq
together with an initial velocity profile $\dot{R}(0,r)$ [which fixes $k(r)$, via Eq.(\ref{tb1}) evaluated at $t=0$].

Without loss of generality, we consider here the $k=0$ case (corresponding to gravitationally unbound matter configurations), since it allows for an analytical solution of Eq. (\ref{tb1}) in closed form:
\bqa
R(t,r)&=&\left(\fr{6m}{\Lambda}\right)^{\fr{1}{3}}\sinh^{2/3}T(t,r), \label{rad} \\
T(t,r)&\equiv&\fr{\sqrt{3\Lambda}}{2}\left[t_{\rm c}(r)-t\right], \label{time}
\eqa
where $t_{\rm c}(r)$ is the proper time for complete collapse [$R(t_{\rm c},r)=0$] of a shell with initial area radius $R(0,r)$, which is fixed by Eq. (\ref{rad}) at $t=0$:
\beq
t_{\rm c}=\fr{2}{\sqrt{3\Lambda}}\sinh^{-1}\left(\sqrt{\fr{\Lambda r^{3}}{6m}}\right), \label{tcoll}
\eeq
where the scaling $R(0,r)=r$ was adopted [note that, when $k=0$, $\dot{R}(0,r)$ is automatically fixed by the choice for the radial gauge via Eq. (\ref{tb1})]. The relevant derivatives of the area radius are
\bqa
R'(t,r)&=&R\left(\fr{m'}{3m}+\sqrt{\fr{\Lambda}{3}}t'_{\rm c}\coth T\right), \label{Rpr} \\
\dot{R}(t,r)&=&-\sqrt{\fr{\Lambda}{3}}R\coth T,
\eqa
where the minus sign corresponds to implosion.

\section{EXISTENCE AND GENERICITY OF SINGULARITIES}

From Eq. (\ref{tb1}), we have $\dot{R}=\pm\sqrt{2mR^{-1}+(\Lambda/3)R^{2}}$, where the plus or minus sign corresponds to expansion or implosion, respectively. If $\dot{R}<0$, then every dust shell implodes and inevitably collapses to vanishing proper area in a proper time given by Eq. (\ref{tcoll}). If $\dot{R}\geq0$, provided the initial ``acceleration'', $\ddot{R}(0,r)=-\fr{m(r)}{r^{2}}+\fr{\Lambda}{3}r<0$, all the shells will initially expand towards increasing area radius, reach a maximum value $R_{\rm max}(r)$, and then collapse back through their original radii, eventually ending up with zero proper area. In either case we have $R(t_{c}(r),r)=0$.

It then follows that at $t=t_{\rm c}(r)$ the Kretschmann scalar,
\beq
{\mathcal K}\equiv R_{abcd}R^{abcd}=3\fr{m'}{R^{4}R'^{2}}-8\fr{mm'}{R^{5}R'}+12\fr{m'^{2}}{R^{6}},
\eeq
diverges, thereby signaling the existence of a curvature singularity. It has been shown (explicitly for the $\Lambda=0$ case, but the result also holds for $\Lambda>0$) that the curvature singularity for events with $r>0$ is spacelike, and thus physically uninteresting \cite{christodoulou84}. Of potential interest is the central curvature singularity that forms at the event $(t=t_{\rm c}(0),r=0)$. 

We want to determine if such a singularity is (at least locally) naked, i.e., if there exists at least one future-oriented radial null geodesic with past endpoint at the singularity. To do so, we examine the outgoing radial null geodesics equation:
\beq
\fr{dt}{dr}=R'=R\left(\fr{m'}{3m}+\sqrt{\fr{\Lambda}{3}}t'_{\rm c}\coth T\right), \label{ng}
\eeq
where Eqs. (\ref{tbm}) and (\ref{Rpr}) were used. Expanding $\rho(0,r)\equiv \rho_{\rm c}(r)$  near $r=0$,
\beq
\rho_{\rm c}(r)=\sum_{i=0}^{+\infty} \rho_{i}r^{i},
\eeq
near the singularity (where $r,T\rightarrow0^{+}$) we have, to leading order,
\bqa
m(r)&=&m_{0}r^{3}+m_{n}r^{n+3}+{\mathcal O}(r^{n+4}), \\
t_{\rm c}(r)&=&t_{0}+t_{n}r^{n}+{\mathcal O}(r^{n+1}), \\
R(t,r)&=&\left(\fr{9}{2}\right)^{\fr{1}{3}}\left(m_{0}^{\fr{1}{3}}r+M_{n}r^{n+1}\right) \left(t_{0}+t_{n}r^{n}-t\right)^{\fr{2}{3}} \nonumber \\
&&+\;{\mathcal O}(r^{n+2})\times{\mathcal O}(T^{\fr{8}{3}}), \label{bigr}
\eqa
where $t_{n}$ and $M_{n}$ are real coefficients linear in $m_{n}=(4\pi/n)\rho_{n}$, with $n>0$; $\rho_{n}\equiv(\partial^{n}\rho_{c}/\partial r^{n})_{r=0}$ is the first non-vanishing derivative of the central energy density distribution, and
\beq
t_{0}=\sinh^{-1}\left(\sqrt{\fr{\Lambda}{6m_{0}}}\right).
\eeq

We now follow the method outlined by Barve {\em et al.} \cite{barveetal99}, and {\em assume} that there is a regular solution to the outgoing null geodesics equation near $r=0$. To leading order in $r$, we write such a solution as
\beq
t=t_{0}+ar^{\sigma}, \label{null}
\eeq
where $a,\sigma\in\Bbb{R}^{+}$. We note that, since $T\geq0$, we must require $\sigma\geq n$. If $\sigma=n$, we have the additional constraint, $a<t_{n}$.

From Eqs. (\ref{bigr}) and (\ref{null}), we obtain
\beq
R(t,r)=\left(\fr{9m_{0}}{2}\right)^{\fr{1}{3}}t_{n}^{\fr{2}{3}}r^{\fr{2n}{3}+1}+{\mathcal O}(r^{\sigma+2-\fr{n}{3}}). \label{newr}
\eeq

Let us first consider the $\sigma>n$ case. Differentiating Eqs. (\ref{null}) and (\ref{newr}) w.r.t. $r$, we obtain, from Eq. (\ref{ng}), to leading order,
\beq
a\sigma r^{\sigma-1}=\left(\fr{2n}{3}+1\right)\left(\fr{9m_{0}}{2}\right)^{\fr{1}{3}}t_{n}^{\fr{2}{3}}r^{\fr{2n}{3}}.
\eeq
Self-consistency fixes 
\bqa
\sigma&=&1+\fr{2n}{3}, \label{slope} \\
a&=&\left(\fr{9m_{0}}{2}\right)^{\fr{1}{3}}t_{n}^{\fr{2}{3}}. 
\eqa
The condition $\sigma>n$ now reads $n<3$.  For $n=1,2$ (i.e., for $\rho_{1}\neq0$, or $\rho_{1}=0$ and $\rho_{2}\neq0$) there is a self-consistent solution to the outgoing radial null geodesics equation in the limit $t\rightarrow t_{0}$, $r\rightarrow0$, and thus there is at least one outgoing radial null geodesic starting from the singularity, which is therefore naked.

We now examine the case $n=\sigma=3$. Proceeding as before, we obtain, to leading order,
\beq
3ar^{2}=3\left(\fr{9m_{0}}{2}\right)^{\fr{1}{3}}(t_{3}-a)^{\fr{2}{3}}r^{2},
\eeq
which is identically satisfied provided
\beq
a^{3}-M^{3}a^{2}+2M^{2}t_{3}a=0, \label{cub}
\eeq
where $M\equiv(9m_{0}/2)^{1/3}$. This equation has two non-zero distinct roots (other than the $a=0$ trivial root), given by $a=(M^{2}/2)\pm\sqrt{M^{4}-8t_{3}}$, if $t_{3}<\fr{1}{8}\left(\fr{9m_{0}}{2}\right)^{\fr{4}{3}}$, which imposes a constraint on $\rho_{3}$, for a given $\rho_{0}$. In addition, self-consistency also requires that $a<t_{3}$, which leads to
\beq
\fr{M^{2}}{2}-4+\sqrt{M^{4}+16-4M^{2}}<t_{3}<\fr{1}{8}M^{4}. \label{t3c}
\eeq
Thus, as long as one restricts ourselves to initial data that satisfies the above condition, the singularity is naked. We note, however, that the $n=3$ case is a less generic case than the $n<3$ one, as it requires that $\rho_{1}=\rho_{2}=0$ and $\rho_{3}$ obey condition (\ref{t3c}).

Let us now investigate whether there is only one null geodesic emanating from the singularity (in which case the singularity would be ``visible'' for an infinitesimal amount of time), or an entire family---in which case the singularity would be visible for an infinite amount of time. We write the equation for the outgoing radial null geodesics to next order as
\beq
t=t_{0}+ar^{\sigma}+br^{\sigma+\delta}, \label{null2}
\eeq
where $a,\sigma,\delta\in\Bbb{R}^{+}$. Again, we shall consider first the case $\sigma<n$. Proceeding as before, we have
\beq
a\sigma r^{\sigma-1}+b(\sigma+\delta)r^{\sigma+\delta-1}=Ar^{\fr{2}{3}\sigma}+Br^{\delta+\fr{2}{3}\sigma},
\eeq
where
\bqa
A&\equiv&\left(\fr{2n}{3}+1\right)\left(\fr{9m_{0}}{2}\right)^{\fr{1}{3}}t_{n}^{\fr{2}{3}}, \\
B&\equiv&\fr{2}{3}\left(\fr{n}{3}-\sigma-1\right)\left(\fr{9m_{0}}{2}\right)^{\fr{1}{3}}t_{n}^{-\fr{1}{3}}.
\eqa
Self-consistency fixes $a$ and $\sigma$ as before, and implies
\beq
b\left(\fr{2n}{3}+1+\delta\right)r^{\delta-1}=Br^{-\fr{n}{3}},
\eeq
thus fixing $\delta=1-\fr{n}{3}$, and $b=-\fr{2}{3}\left(\fr{9m_{0}}{2}\right)^{\fr{1}{3}}t_{n}^{-\fr{1}{3}}$. Hence, to this order there is a single future-oriented radial null geodesic emanating from the singularity.

Let us now examine the case $\sigma=n=3$. From Eqs. (\ref{bigr}) and (\ref{null2}), we get
\beq
R(t,r)=\alpha_{1}r^{3}+\alpha_{2}r^{3+\delta},
\eeq
where
\bqa
\alpha_{1}&\equiv&\left(\fr{9m_{0}}{2}\right)^{\fr{1}{3}}(t_{3}-a)^{\fr{2}{3}}, \\
\alpha_{2}&\equiv&-\fr{2}{3}\alpha_{1}(t_{3}-a)^{-1}.
\eqa

Taking the partial derivative of $R(t,r)$ w.r.t. $r$ and equating it to that of Eq. (\ref{null2}) yields
\beq
3ar^{2}+b(3+\delta)r^{2+\delta}=3\alpha_{1}r^{2}+\alpha_{2}(3+\delta)br^{2+\delta}.
\eeq
Self-consistency at first-order fixes $a=\alpha_{1}$ [which is equivalent to the cubic equation (\ref{cub})] as before, and a solution exists if condition (\ref{t3c}) is satisfied. At the next order, $b$ drops out and self-consistency requires $\alpha_{2}=1$. This amounts to a {\em particular} value of $a$, which is consistent with those determined from Eq. (\ref{cub}), if and only if
\beq
t_{3}=M^{3}+\fr{M^{2}}{2}-4\pm\sqrt{M^{4}-8M^{3}+4M^{2}+4}. \label{newt3}
\eeq
Positivity of the radicand requires $M<1.076381196$, or $M>7.453694050$, but the first range is disallowed since we must have $t_{3}>0$. Consistency between condition (\ref{t3c}) and Eq. (\ref{newt3}) further imposes $M>8.799799016$. Hence, a fully self-consistent solution with $n=3$ exists when $M>8.799799016$ {\em and} $t_{3}$ is given by Eq. (\ref{newt3}). In such case, an entire one-parameter family of outgoing radial null geodesics (parameterized by $b$) departs from the singularity. Clearly, this is a very special case---it is a set of measure zero in the initial data\footnote{We note that, while this is obvious in the present case, it is in general unclear what measure or topology should be imposed on the space of initial data.}---and in general there is only a single outgoing null geodesic with past endpoint at the singularity.

\section{GLOBAL VISIBILITY}

We have shown that there is a generic central curvature singularity in Tolman-Bondi-de Sitter collapse. This singularity is at least locally naked. It would be globally naked if outgoing geodesics starting from the central singularity at $(t=t_{0},r=0)$ could reach future null infinity. In the present case, this would require future-directed geodesics to avoid the event horizon (i.e., remain outside it at all times) and cross the cosmological horizon.

Let us first examine the possibility of event horizon avoidance. In spherical dust collapse, the event horizon (EH) coincides with the apparent horizon (AH) at the boundary of the spherical mass distribution, $r=r_{\rm m}$. Hence, in order for the outgoing null geodesics to escape the EH, they must cross $r=r_{\rm m}$ {\em before} the AH, thereby avoiding becoming trapped, and hence eventually ingoing. 

The AH is a spacelike 2-surface defined by the locus of events where null wavefronts become ``frozen'', and in the adopted coordinates it is given by $R_{,a}R_{,b}g^{ab}=0$. With the metric (\ref{tbm}) and Eq. (\ref{tb1}) we have then
\beq
\fr{\Lambda}{3}R^{3}-R+2m=0.
\eeq
This equation has three distinct real roots if $3m\sqrt{\Lambda}<1$, two of which are positive and given by
\bqa
R_{1}&=&\fr{2}{\sqrt{\Lambda}}\sin\left[\fr{1}{3}\sin^{-1}(3m\sqrt{\Lambda})\right], \label{rbh} \\
R_{2}&=&\fr{2}{\sqrt{\Lambda}}\sin\left[\fr{1}{3}\sin^{-1}(3m\sqrt{\Lambda})+\fr{2\pi}{3}\right], \label{rcsm}
\eqa
with $R_{2}>R_{1}>0$, corresponding to the choice $0\leq\sin^{-1}\beta\leq\pi/2$, $0\leq\beta\leq1$. The third root, $R_{3}=-R_{1}-R_{2}$, is negative and hence unphysical. $R_{2}$ is a generalized cosmological horizon ($R_{2}=\sqrt{3/\Lambda}$, when $m=0$) and $R_{1}$ the black hole apparent horizon ($R_{1}=2m$ when $\Lambda=0$; the apparent and event horizons coincide in the static case). For $3m\sqrt{\Lambda}=1$, the two horizons coincide. If $3m\sqrt{\Lambda}>1$, there is one negative real root and two complex (conjugate) roots, all of which are unphysical: the spacetime does not admit any horizons in this case.

From Eq. (\ref{rbh}) together with Eqs. (\ref{rad})-(\ref{time}), we obtain
\beq
t_{\rm AH}(r)=t_{\rm c}(r)-\fr{p}{\sqrt{3}}\sinh^{-1}\left[\fr{2}{p}\sqrt{\fr{1}{\alpha}}\left(p\sin\Theta\right)^{\fr{3}{2}}\right], \label{tah}
\eeq
where
\bqa
p&\equiv&\fr{2}{\sqrt{\Lambda}}, \\
\alpha&\equiv&\fr{6m}{p}, \\
\Theta&\equiv&\fr{1}{3}\sin^{-1}\alpha.
\eqa
A necessary condition for the singularity to be globally naked is $t_{\rm AH}(0)>t_{\rm c}(0)$, which ensures that radial null geodesics emanating from the singularity do so before the AH forms. However, from Eq. (\ref{tah}) we have $t_{\rm AH}(0)=t_{\rm c}(0)-X$, where $0\leq X<(2/\sqrt{3\Lambda})\sinh^{-1}[(\sqrt{3}/m)^{1/2}/\Lambda]$. Outgoing radial null geodesics are therefore unavoidably trapped inside the AH and the singularity cannot be globally naked.

\begin{figure}
\begin{center}
\epsfxsize=15pc
\epsffile{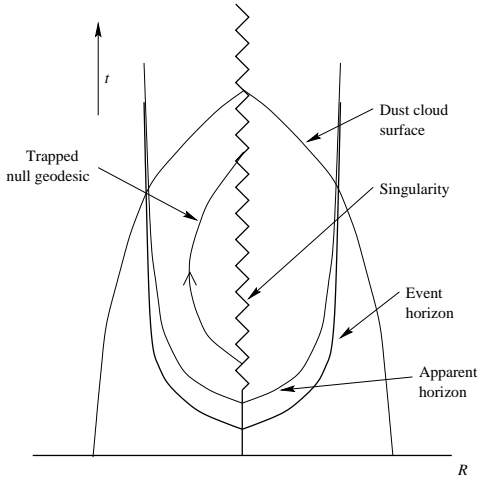}
\end{center}
\caption{Central naked singularity in spherical Tolman-Bondi-de Sitter collapse. The singularity is locally naked: it forms before the boundary of the mass distribution undergoes complete collapse, but after the apparent (and thus event) horizon form. Null geodesics emanating from the singularity are necessarily trapped.\label{fig1}}
\end{figure}

\section{CURVATURE STRENGTH}

A crucial property of a singularity is its curvature strength. A singularity is said to be gravitationally strong in the sense of Tipler \cite{tipler77} if every collapsing volume element is crushed to zero at the singularity, and weak otherwise (i.e., if it remains finite). It is generally believed---although not yet proven \cite{clarke93}---that spacetime is geodesically incomplete at a strong singularity, but extendible through a weak one \cite{tipler77,tipler&clarke&ellis80,ori92}.

A precise characterization of Tipler strong singularities has been given by Clarke and Kr\' olak \cite{clarke&krolak85}, who proposed (among other conditions) the {\em strong focusing condition}: There is at least one null geodesic, with tangent $k^{a}$ and affine parameter $\lambda$ (with $\lambda=0$ at the singularity), along which the following is satisfied:
\beq
\lim_{\lambda\rightarrow0} \lambda^{2}R_{ab}k^{a}k^{b}>0. \label{strong}
\eeq
This is a sufficient condition for the singularity to be Tipler strong and corresponds to the vanishing of any two-form defined along such a geodesic, at the singularity, due to unbounded curvature growth.

Let us now consider a radial null geodesic with tangent $k^{a}=(k^{t},k^{r},0,0)$, where $k^{a}=\fr{dx^{a}}{d\lambda}$, and 
\beq
k^{t}\equiv FR=R'k^{r}, \label{kt}
\eeq
where $F$ can be written as an explicit function of the affine parameter, $F=F(\lambda)$, obeying the differential equation (which follows from the geodesic equation, $k^{a}\nabla_{a}k^{b}=0$):
\beq
\fr{dF}{d\lambda}+F^{2}\left(1+\dot{R}+R\fr{\dot{R}'}{R'}\right)=0. \label{dfdl}
\eeq

From Eqs. (\ref{tbm}) and (\ref{tb1}) we have then
\beq
\Omega\equiv R_{ab}k^{a}k^{b}=2(k^{t})^{2}\fr{1}{R}\left(\fr{\dot{R}\dot{R}'}{R'}-\ddot{R}\right)=2(k^{t})^{2}\fr{m'}{R^{2}R'}.
\eeq
Using Eqs. (\ref{rad})-(\ref{null}) and (\ref{kt})-(\ref{dfdl}), we get
\bqa
\lim_{\lambda\rightarrow0} \lambda^{2}\Omega&=&\lim_{\lambda\rightarrow0} \lambda^{2}F^{2}\fr{m'}{R'}=\lim_{t\rightarrow t_{0},r\rightarrow0}\chi^{2}\fr{m'}{R'} \nonumber \\
&=&K_{n}\lim_{t\rightarrow t_{0},r\rightarrow0}\chi^{2},
\eqa
where l'H\^ opital's rule was used twice in the second equality, and
\bqa
\chi(t,r)&\equiv&-\left(1+\dot{R}+R\fr{\dot{R}'}{R'}\right)^{-1} \\
&=&-\left(1-\fr{\fr{4m}{R}-\fr{\Lambda}{3}R^{2}-2\fr{m'}{R'}}{\sqrt{\fr{2m}{R}+\fr{\Lambda}{3}R^{2}}}\right)^{-1},
\eqa
where Eq. (\ref{tb1}) was used, and
\bqa
K_{n}&\equiv&\lim_{t\rightarrow t_{0},r\rightarrow0}\fr{m'}{R'}=\fr{1}{3}\left(\fr{2n}{3}+1\right)\lim_{t\rightarrow t_{0},r\rightarrow0}\fr{m}{R} \nonumber \\
&=&\bar{K}\lim_{r\rightarrow0} r^{2-\fr{2n}{3}}=0, \;\;\;\;\;\;\;\;\;\;\;\;\;\;\;\;\;\;\mbox{for $n<3$}, \\
K_{3}&=&\left(\fr{2m_{0}^{2}}{9}\right)^{\fr{1}{3}}\left(t_{3}-a\right)^{-\fr{2}{3}}>0, \;\;\;\;\;\mbox{for $n=3$}, \\
\bar{K}&\equiv&3\left(\fr{2n}{3}+1\right)^{-1}\left(6m_{0}^{2}\right)^{\fr{1}{3}}t_{n}^{-\fr{2}{3}}, \;\;\mbox{for $n<3$}.
\eqa
[Note that $K_{3}$ exists as long as the initial data obeys condition (\ref{t3c})].

Therefore, for $n<3$,
\beq
\lim_{t\rightarrow t_{0},r\rightarrow0} \chi^{2}=\lim_{r\rightarrow0} \left(1-f(n)\sqrt{\bar{K}}r^{3-n}\right)^{-2}=1,
\eeq
where $f(n)\equiv[(8n/3)-(2/3)]/\sqrt{2[(2n/3)+1]/3}>0$. 

If $n=3$, we have
\beq
\lim_{t\rightarrow t_{0},r\rightarrow0} \chi^{2}=(1-\sqrt{2K_{3}})^{-2}\equiv\chi_{0}^{2},
\eeq
which is positive definite provided $K_{3}\neq1/2$.

Summarizing, the singularity is gravitationally strong in the sense of Tipler if $n=3$. For $n<3$, $\lim_{\lambda\rightarrow0} \lambda^{2}R_{ab}k^{a}k^{b}=0$, and the singularity may or may not be Tipler strong.

\subsection{Weyl curvature}

In order to gain further insight on the nature of the central singularity, we examine the behavior of the Weyl curvature scalar in the limit $t\rightarrow t_{0}$ at $r=0$. In a four-dimensional manifold, the Weyl tensor is defined as
\beq
C_{abcd}=R_{abcd}+g_{a[d}R_{c]b}+g_{b[c}R_{d]a}+\fr{1}{3}Rg_{a[c}g_{d]b},
\eeq
and represents the part of Riemannian curvature which is not locally determined by the matter distribution. A divergence of the Weyl scalar, $C\equiv C_{abcd}C^{abcd}$, indicates a blow-up in curvature caused not by the local matter distribution---as in the case of the Ricci scalar, or any other scalar constructed solely from the Ricci tensor---but by the matter content of the spacetime at other points. 

In the metric (\ref{tbm}), the non-vanishing independent components of the Weyl tensor are
\bqa
C_{trtr}&=&2R'^{2}\left(\fr{m'}{R^{2}R'}-\fr{m}{R^{3}}\right), \\
C_{t\theta t\theta}&=&\fr{m}{R}-\fr{m'}{3R'}, \\
C_{t\phi t\phi}&=&\sin^{2}\theta C_{t\theta t\theta}, \\
C_{r\theta r\theta}&=&\fr{m'R'}{3}-\fr{mR'^{2}}{R}, \\
C_{r\phi r\phi}&=&\sin^{2}\theta C_{r\theta r\theta}, \\
C_{\theta\phi\theta\phi}&=&2\sin^{2}\theta\left(mR-\fr{m'R^{2}}{3R'}\right).
\eqa
The Weyl scalar is
\beq
C(t,r)=\fr{48}{R^{4}}\left(\fr{m'}{3R'}-\fr{m}{R}\right)^{2}.
\eeq
At the singularity we obtain, for $n<3$,
\beq
C_{\rm sing}=\lim_{t\rightarrow t_{0}, r\rightarrow 0} C(t,r)=C_{0}\lim_{r\rightarrow0} r^{-\fr{16n}{3}}=+\infty,
\eeq
where $C_{0}\equiv48(2\bar{K}n/9)^2(9m_{0}/2)^{-4/3}t_{n}^{-8/3}>0$. 

If $n=3$,
\beq
C_{\rm sing}=\lim_{r\rightarrow0} 48\left(\fr{9m_{0}}{3}\right)^{2}t_{3}^{-\fr{8}{3}}K_{3}^{2}r^{-12}=+\infty.
\eeq
Thus, for all the initial data leading to naked singularities ($n\leq3$), the Weyl scalar diverges along outgoing null geodesics at the singularity. We note that the present analysis, for outgoing radial null geodesics, remains unchanged up to a sign for $dt/dr$ and $a$ in Eq. (\ref{ng}), for ingoing radial null geodesics. We conclude, therefore, that the Weyl curvature scalar diverges at the singularity along both outgoing {\em and} ingoing null geodesics. This is in agreement with the results of Barve and Singh \cite{barve&singh97} for asymptotically flat  spherical dust collapse, and goes against the speculation by Penrose \cite{penrose81} that the Weyl curvature should diverge along ingoing geodesics, and vanish along outgoing geodesics terminating at the singularity.

Finally, we compute the Ricci curvature scalar, ${\mathcal R}$, at the singularity and compare it to the Weyl scalar, to determine the relative contributions of curvature (i.e., locally versus non-locally induced). The Ricci scalar is
\beq
{\mathcal R}(t,r)=\fr{2m'}{R^{2}R'}+4\Lambda.
\eeq
At the singularity we have, for $n<3$,
\beq
{\mathcal R}_{\rm sing}=\lim_{t\rightarrow t_{0}, r\rightarrow 0} {\mathcal R}(t,r)={\mathcal R}_{0}\lim_{r\rightarrow0} r^{-2n}=+\infty,
\eeq
where ${\mathcal R}_{0}\equiv2(9m_{0}/2)^{-2/3}t_{n}^{-4/3}>0$.

For $n=3$
\beq
{\mathcal R}_{\rm sing}=\lim_{r\rightarrow0} 2K_{3}\left(\fr{9m_{0}}{2}\right)^{-\fr{2}{3}}t_{3}^{-\fr{4}{3}}r^{-6}=+\infty.
\eeq
Hence, $(C/{\mathcal R})_{\rm sing}$ diverges as $r^{-10n/3}$, for $n<3$, and as $r^{-6}$ for $n=3$. Such a dominance of the Weyl curvature over Ricci indicates a predominantly non-local origin for the Riemannian curvature divergence at the singularity. 

\section{CONCLUSIONS}

We have examined in detail the central curvature singularity occurring in general inhomogeneous spherical dust collapse in an asymptotically de Sitter spacetime. This singularity was found to be locally naked for a wide class of initial data: $\rho_{1}\neq0$, or $\rho_{1}=0$ and $\rho_{2}\neq0$, or $\rho_{1}=\rho_{2}=0$ and $\rho_{3}\neq0$. In the latter case (if, in addition, $\rho_{3}$ is determined by $\rho_{0}$), an entire one-parameter family of outgoing radial null geodesics has its past endpoint at the singularity, which is thence visible for a finite amount of time. One must note, however, that this case is non-generic, whereas the first two are generic insofar as initial data is concerned. Hence, generically, a single null geodesic escapes the singularity.

Irrespective of the initial data, the singularity cannot be globally naked, since it forms after the apparent horizon does, and, consequently, any null geodesics emanating from it are necessarily trapped and thus cannot escape the event horizon.

Regarding curvature strength, our analysis, based on the sufficient condition by Clarke and Kr\' olak \cite{clarke&krolak85}, revealed that the singularity is Tipler strong for the case $n=3$ and may or may not be Tipler strong if $n<3$. We note that, while more restrictive than the cases $n<3$, the $n=3$ case is of finite measure in the space of initial data---$\rho_{3}$ has a finite real range, given implicitly by condition (\ref{t3c})---and hence generic, in the sense that it has codimension two in the (countable, infinite-dimensional) space of derivatives of $\rho_{\rm c}(r)$.

We also found that both Weyl and Ricci scalars diverge at the singularity along ingoing and outgoing geodesics. The Weyl scalar divergence dominates over the Ricci, indicating a predominantly non-local origin of the Riemannian curvature unbounded growth. Interestingly, while features such as global visibility and curvature strength appear to depend critically on the local matter distribution, the divergence of Riemann curvature invariants (e.g., Kretschmann scalar) seems to be associated with the matter distribution at other points. It is tempting to speculate that one could alter the local matter distribution, so as to cure such ``problems'' as local visibility and Tipler strong strength, while still maintaining a divergent Riemannian curvature invariant (built solely from the Riemann tensor, without any of its internal index contractions). This naturally leads to the question of whether spacetime can be extendible through a Tipler weak singularity where curvature is, nevertheless, divergent.

Finally, in the context of cosmic censorship, with the current status of $\Lambda$, any realistic formulation (of the strong version, at least) {\em might} have to exclude the cases corresponding to $\Lambda>0$, {\em if} a stability analysis---to be defined in a suitable, precise manner---would reveal the singularity to be a persistent (i.e., exist for all times, with the same properties) feature of the spacetime. The results of Deshingkar, Joshi, and Dwivedi \cite{deshingkar&joshi&dwivedi99}, and those of Harada, Iguchi, and Nakao \cite{hin98-00}, reveal that the central singularity in spherical dust collapse with $\Lambda=0$ is stable (i.e., remains locally naked and Tipler strong) against initial data perturbations, and marginally stable against metric (and matter {\em coupled} to metric) perturbations. The similarity between such a singularity and the one discussed in the present paper---the conditions for visibility and the slope of the outgoing radial null geodesics [cf. Eq. (\ref{slope})] with past endpoint at the singularity are the {\em same} in both cases (see, e.g., \cite{barveetal99})---suggests that analogous stability properties may also hold for the central curvature singularity in Tolman-Bondi-de Sitter collapse. This issue is currently under investigation \cite{goncalves00b}. 

\section{ACKNOWLEDGMENTS}
I am grateful to Kip Thorne for useful discussions. This work was supported by FCT (Portugal) Grant PRAXIS XXI-BPD-16301-98, and by NSF Grant AST-9731698.

\end{document}